\documentclass[prl, aps, floatfix, superscriptaddress, preprint, 12pt]{revtex4-2}
\usepackage[a4paper, total={170mm,257mm}]{geometry}
\usepackage{graphicx}
\usepackage{amsmath}
\usepackage{xcolor}
\usepackage{layouts}

\begin{document}

\title{Breakdown of the critical state in the ferromagnetic superconductor EuFe$_2$(As$_{1-x}$P$_x$)$_2$}

\author{William Robert Fern}
\thanks{These authors contributed equally to this work.}
\affiliation{Department of Physics, University of Bath, Claverton Down, Bath, BA2 7AY, United Kingdom}

\author{Joseph Alec Wilcox}
\thanks{These authors contributed equally to this work.}
\email{Corresponding author: jaw73@bath.ac.uk}
\affiliation{Department of Physics, University of Bath, Claverton Down, Bath, BA2 7AY, United Kingdom}

\author{Tong Ren}
\affiliation{Department of Applied Physics, The University of Tokyo, 7-3-1 Hongo, Bunkyo-ku, Tokyo 113-8565, Japan}

\author{Ivan Veshchunov}
\affiliation{Department of Applied Physics, The University of Tokyo, 7-3-1 Hongo, Bunkyo-ku, Tokyo 113-8565, Japan}

\author{Tsuyoshi Tamegai}
\affiliation{Department of Applied Physics, The University of Tokyo, 7-3-1 Hongo, Bunkyo-ku, Tokyo 113-8565, Japan}

\author{Simon John Bending}
\affiliation{Department of Physics, University of Bath, Claverton Down, Bath, BA2 7AY, United Kingdom}

\begin{abstract}

There are very few materials in which ferromagnetism coexists with superconductivity due to the destructive effect of the magnetic exchange field on singlet Cooper pairs. The iron-based superconductor EuFe$_2$(As$_{1-x}$P$_x$)$_2$ is therefore unique in exhibiting robust superconductivity with a maximum critical temperature of 25~K and long-range ferromagnetism below $T_\mathrm{FM}\approx19$~K. Here we report a spatially-resolved study of the irreversible magnetisation in this system that reveals a variety of novel behaviours that are strongly linked with underlying ferromagnetic domain structures. In the superconducting-only state, hysteretic magnetisation due to irreversible vortex motion is consistent with typical weak vortex-pinning behaviour. Just below $T_\mathrm{FM}$, very narrowly-spaced stripe domains give rise to highly erratic and irreproducible fluctuations in the irreversible magnetisation that we attribute to the dynamics of multi-vortex clusters stabilised by the formation of vortex polarons. In contrast, at lower temperatures, ferromagnetic domains become wider and saturated with spontaneously nucleated vortices and antivortices, leading to a smoother but unconventional evolution of the irreversible state. This observation suggests that the penetrating flux front is roughened by the presence of the magnetic domains in this regime, presenting a clear departure from standard critical state models. Our findings indicate that the mechanism governing irreversibility is strongly influenced by the precise nature of the underlying ferromagnetic domains, being very sensitive to the specific material parameters of EuFe$_2$(As$_{1-x}$P$_x$)$_2$. We consider the possible microscopic origins of these effects, and suggest further ways to explore novel vortex-domain magnetic behaviours.

\end{abstract}

\maketitle

\section{Introduction}

The coexistence of ferromagnetism and superconductivity in a single material is highly unusual due to the typically destructive effect of ferromagnetism on the superconducting order parameter\cite{bulaevskiiCoexistenceSuperconductivityMagnetism1985, liuSuperconductivityFerromagnetism2016}. There are relatively few examples of conventional superconductors where the two coexist freely, as the ferromagnetic exchange field tends to align the spins of singlet Cooper pairs and ultimately destroy superconductivity\cite{matthiasSpinExchangeSuperconductors1958,kulicConventionalMagneticSuperconductors2006,wolowiecConventionalMagnetic2015}. The coexistence of ferromagnetism and superconductivity is slightly more common in unconventional superconductors, such as in some U-containing compounds\cite{aokiFerromagnetismSuperconductivityUranium2012,aokiReviewUbasedFerromagnetic2019} e.g.\ URhGe\cite{aokiCoexistenceSuperconductivityFerromagnetism2001} and UCoGe\cite{paulsenObservationMeissnerOchsenfeldEffect2012}. Here the adverse effect of the ferromagnetic exchange field is avoided by forming triplet Cooper pairs whose spins are parallel with the underlying ferromagnetic order; however, these materials usually exhibit very low superconducting critical temperatures as a result. The coexistence of antiferromagnetism and singlet superconductivity, in contrast, is more frequently observed as the spin polarization of magnetic ions tends to average to zero on the length scale of the superconducting coherence length\cite{wolowiecConventionalMagnetic2015}. Examples of coexisting antiferromagnetism and superconductivity can be found in both conventional superconductors\cite{kulicConventionalMagneticSuperconductors2006}, such as NdRh$_4$B$_4$\cite{mapleSuperconductivityLongRange1980}, and also amongst unconventional heavy-fermion compounds\cite{whiteUnconventionalSuperconductivity2015} e.g.\ CeRhIn$_5$\cite{parkHiddenMagnetism2006}, though once again with low critical temperatures of the order of only a few Kelvin.

The relatively recent discovery of several magnetic, Eu-containing iron-pnictide superconductors, with $T_c$ up to $\sim40$~K\cite{micleaEvidenceReentrant2009,anupamSuperconductivityMagnetism2009,liuSuperconductivityFerromagnetism2016,liuNewFerromagnetic2016}, therefore presents a new paradigm in magnetic superconductors and a rare opportunity to study the coexistence of robust superconductivity with long-range magnetic order. In these materials, the superconductivity originates in the Fe-pnictide planes, while the magnetism stems from the localised Eu$^{2+}$ moments, providing spatial separation between the two in the crystal structure\cite{liuIronbasedMagneticSuperconductors2022}. The relatively low magnetic ordering temperature of the Eu moments, $\sim10$ - 20~K, also gives rise to a weak exchange field such that the influence on the superconductivity is predominantly via electromagnetic coupling rather than the exchange interaction. In EuFe$_2$(As$_{1-x}$P$_x$)$_2$, a dome of superconductivity, with maximum $T_c\sim25$~K at $x\sim0.2$, is induced through the isovalent substitution of P for As\cite{renSuperconductivityInduced2009}, as shown schematically in figure \ref{fig:1}(a). In this same region of the phase diagram, the Eu-based magnetic order also switches from initially antiferromagnetic to ferromagnetic, and with a nearly constant ordering temperature $T_\mathrm{FM}\sim19$~K\cite{caoSuperconductivityFerromagnetism2011}. By careful tuning of the P-composition $x$, it is possible therefore to produce samples with either $T_c>T_\mathrm{FM}$ or $T_c<T_\mathrm{FM}$, and with a large window of coexistence up to $\Delta{T}\approx19$~K.

Recently, magnetic force microscopy (MFM) imaging has revealed a rich variety of unique magnetic phenomena in this system. At the optimal composition $x\sim0.2$, where $T_c>T_\mathrm{FM}$, a very fine uniaxial stripe domain structure appears in a narrow window of temperature below $T_\mathrm{FM}$, known as the domain Meissner state (DMS)\cite{stolyarovDomainMeissnerState2018}. These domains, with size less than the London penetration depth, are magnetised parallel to the crystalline $c$-axis and alternate between `up' and `down' orientations. The superconductivity responds to the presence of the domains by generating screening currents that run near the domain walls, suffering an increase in kinetic energy as a result\cite{devizorovaTheoryMagneticDomain2019}. As the temperature is reduced, the magnetisation of the domains increases which in turn increases the energy cost of the screening currents. Eventually, after only a narrow window of existence $\Delta{T}\sim1$~K, the cost to the superconductivity becomes too great and the screening currents collapse. The system now transitions to a second, distinct magnetic state, the domain vortex state (DVS), where the up and down domains become saturated with spontaneously nucleated superconducting vortices and antivortices respectively. The loss of the screening currents also causes the domains to recover their much larger `natural' width compared to the DMS.

A more recent study of vortex behaviour in large magnetic fields demonstrated irreversible, global magnetic behaviour that was clearly correlated with the presence of the two distinct ferromagnetic domain structures. In particular, a novel phenomenon was identified as resulting from the interaction of superconducting vortices with the fine ferromagnetic stripe domains\cite{wilcoxMagneticallyControlledVortex2025}, where the magnetic field associated with a superconducting vortex causes a local perturbation of the surrounding domain widths - collectively termed a vortex polaron. Due to their lower energy compared to a free vortex, an increase in the critical current and vortex pinning potential were both attributed to the presence of vortex polarons. Surprisingly, vortex polarons also exhibit a short-range, attractive potential; MFM imaging demonstrated conclusively the formation of closely-bound pairs and chains of multiple vortices that were capable of substantially modifying and restructuring the local domains. The influence of vortex polarons was found to be strongest in the DMS, where the ratio of the London penetration depth to the domain width, $\lambda/w_D$, is largest. The expansion of the domains in the DVS, and reduction of $\lambda$ with decreasing temperature, instead leads to a strong reduction in the vortex polaron binding energies  at low temperatures.

We note that coupling of ferromagnetic stripe domain phases with superconductivity has previously observed in a hybrid Nb-film / Fe-garnet bilayer structure\cite{vlasko-vlasovCoupledDomainStructures2010,vlasko-vlasovDomainStructureMagnetic2012}, resembling the DMS of EuFe$_2$(As$_{1-x}$P$_x$)$_2$, and the spontaneous nucleation of vortex-antivortex pairs has also been seen in a hybrid Nb-film / Permalloy planar structure, resembling the DVS\cite{iavaroneImagingSpontaneousFormation2011}. However, the width of the magnetic domains reported for these hybrid structures is considerably larger ($\sim$ few $\mu$m) than what was reported for EuFe$_2$(As$_{1-x}$P$_x$)$_2$ ($\sim$ hundreds of nm)\cite{stolyarovDomainMeissnerState2018,wilcoxMagneticallyControlledVortex2025}. It is important to emphasise that the small width of the ferromagnetic domains compared to the vortex size is critical to driving the vortex polaron behaviour in EuFe$_2$(As$_{1-x}$P$_x$)$_2$\cite{wilcoxMagneticallyControlledVortex2025}.

The irreversible behaviour of vortices penetrating type-II superconductors can frequently be understood in terms of the Bean critical state model whereby the spatial gradient of the magnetic induction is limited by the superconducting critical current, $J_c$ \cite{beanMagnetizationHardSuperconductors1962,beanMagnetizationHighFieldSuperconductors1964}. While the model contains many simplifying assumptions, the critical state picture, and subsequent modifications for differing sample geometries, e.g.\ by Brandt and Indembom\cite{brandtTypeIIsuperconductorStripCurrent1993}, has proven to be highly successful in understanding the hysteretic magnetisation behaviour of type-II superconductors\cite{ruizCriticalCurrentDensity2026}. However, the presence of underlying ferromagnetic stripe domains in EuFe$_2$(As$_{1-x}$P$_x$)$_2$, represents a dramatic departure from the model assumption that the vortex pinning force is uniform and spatially homogeneous, as in Bean's simple picture of a type-II superconductor. While highly-local MFM imaging has so far revealed the intimate link between domains and superconducting vortices, as well as their resulting influence on the `bulk' magnetic behaviour\cite{wilcoxMagneticallyControlledVortex2025}, an understanding of how this affects the macroscopic flux distribution that ultimately determines the global irreversibility is currently lacking. Here we present a spatially-resolved study of the irreversible flux penetration and dynamics in optimally and over-doped samples of EuFe$_2$(As$_{1-x}$P$_x$)$_2$. At optimal doping, we find that the hysteretic magnetic behaviour is acutely sensitive to the underlying ferromagnetic state, exhibiting a transition from typical type-II superconducting behaviour above $T_\mathrm{FM}$ to a highly chaotic and irreproducible magnetic response in the DMS that is accompanied by a sharp suppression in the superconducting order parameter. Upon cooling into the DVS, the erratic response abruptly disappears and is replaced by a strikingly smooth behaviour that differs fundamentally from that observed above $T_\mathrm{FM}$ due to the presence of a very high-density background of spontaneously generated flux. The behaviour of the over-doped sample, with $T_c < T_\mathrm{FM}$, is also found to depart significantly from a normal type-II superconductor, exhibiting marked differences from the optimally doped crystal for comparable magnetic states. Our observations suggest that the interplay between the magnetic and superconducting orders in EuFe$_2$(As$_{1-x}$P$_x$)$_2$ depends very sensitively on the specific material parameters, with the precise nature of the underlying ferromagnetic state a key factor governing magnetic irreversibility. 

\section{Methods}

\begin{figure*}
    \includegraphics[width=\linewidth]{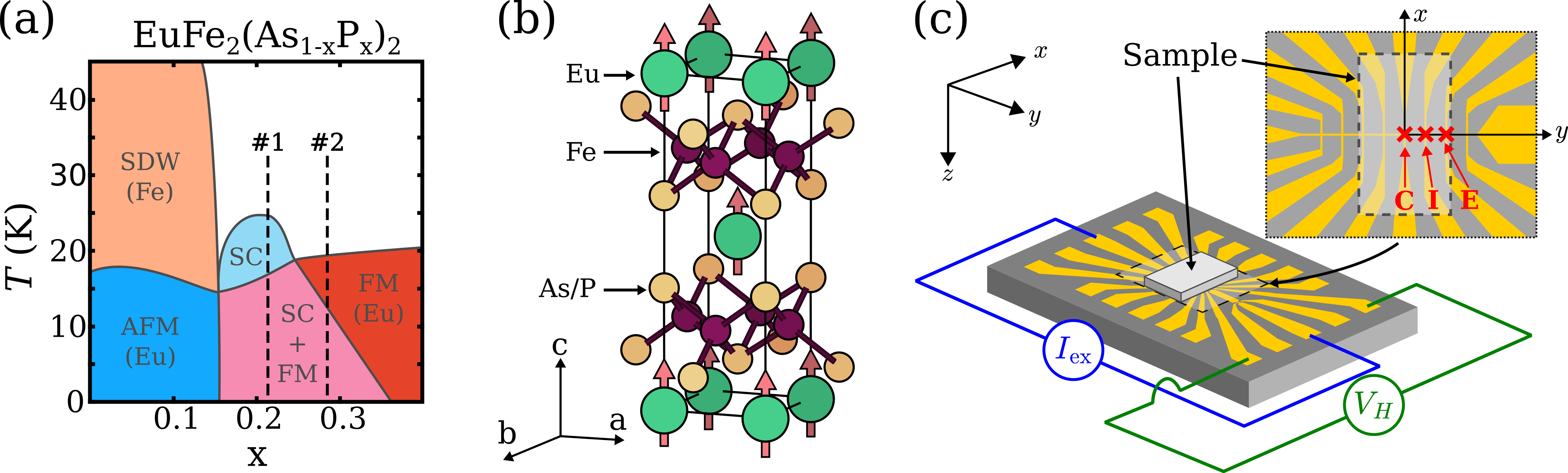}
    \centering
    \caption{(a) Schematic phase diagram of EuFe$_2$(As$_{1-x}$P$_x$)$_2$ with approximate positions of samples \#1 and \#2 indicated. (b) Tetragonal crystal structure of EuFe$_2$(As$_{1-x}$P$_x$)$_2$. The pink arrows indicate the direction of the Eu$^{2+}$ moments in the ferromagnetic state. (c) Schematic of experimental setup with a sample placed on top of a patterned HSA device. The inset shows an expanded, top-down view of the sample on the device, and the nominal positions of the centre (C), intermediate (I) and edge (E) sensors are indicated by the red crosses.}
    \label{fig:1}
\end{figure*}

Here we present a study on the critical state in two platelet-shaped samples of EuFe$_2$(As$_{1-x}$P$_x$)$_2$, \#1 and \#2, with the shortest dimension corresponding to the crystal $c$-axis (see figure \ref{fig:1}(b)). Sample \#1 has nominally optimal P-composition $x\approx0.21$, as indicated in figure \ref{fig:1}(d), with $T_c\approx24$~K~$>T_\mathrm{FM}\approx19.3$~K and dimensions $l~{\times}~w~{\times}~t$~=~$388\,\mu$m$~{\times}~297\,\mu$m$~{\times}~10~\mu$m. Sample \#2, in contrast, has higher P-composition $x\approx0.28$ such that $T_c\approx12$~K~$<T_\mathrm{FM}\approx19.3$~K, and dimensions $l~{\times}~w~{\times}~t$~=~$641\,\mu$m$~{\times}~410\,\mu$m$~{\times}~33~\mu$m. A schematic phase diagram for the system with the approximate positions of samples \#1 and \#2 is shown in figure \ref{fig:1}(a). Details of crystal growth are provided in \textit{Supplementary Note 1}.

We probe the magnetic profile of our samples by utilising devices containing a linear array of microscopic Hall sensors (Hall sensor array, HSA). These devices are fabricated from a GaAs/AlGaAs heterostructure containing a two-dimensional electron gas (2DEG) located $\sim100$ nm below the surface. The 2DEG has a (dark) $T=1.5$~K carrier concentration $n_e=1.6\times10^{11}$~cm$^{-2}$ and electron mobility $\mu\approx1.5\times10^6$ cm$^2$~V$^{-1}$~s$^{-1}$. Using photolithography, the 2DEG is patterned into a linear array of nine Hall sensors. The sensors share a common current channel, have an active area of $5\times5$~$\mu$m$^2$, and centre-to-centre separation of 50~$\mu$m; a schematic is shown in figure \ref{fig:1}(c). Of the nine sensors, one sensor is located further away at a distance of 650~$\mu$m from the nearest sensor (the edge sensor, E) in order to provide a reference measurement of the applied magnetic field without influence from the samples' magnetic response (see \textit{Supplementary Note 2} for further details).

The samples are mounted on the surface of the HSA devices using a low melting-temperature paraffin wax, positioned symmetrically across the current channel and with the shorter sample edges parallel to the current channel, as shown in figure \ref{fig:1}(c). Sample \#1 was positioned such that three sequential sensors were positioned at $y$~=~$8,\,58,\,108\,\mu$m from the sample centre. Due to its larger dimensions, alternate sensors with positions $y$~=~$-4,\,96,\,196\,\mu$m were used for sample \#2. These three different sensor positions we will refer to as centre (C), intermediate (I) and edge (E), respectively.

An applied magnetic field $H_a$ is created using a custom copper-wound short solenoid oriented with its central axis normal to the plane of the device (i.e.\ along the $z$-direction) and thus parallel to the $c$-axis of the samples, as shown in figure \ref{fig:1}(c). The use of a resistive electromagnet ensures the remanent field is as small as possible when the samples are cooled through $T_c$. A 37.0 Hz AC excitation current $I_\mathrm{ex}=5\mu$A is applied along the central channel of the HSA and the transverse voltage $V_H$ that develops across the individual sensors is measured using a lock-in amplifier. In the absence of a sample, such that $B=\mu_0H_a$, the transverse voltage is due to the Hall coefficient $R_H$ of a particular sensor (see \textit{Supplementary Note 2}) and the component of the magnetic induction in the $z$-direction, $B_z$:
\begin{equation}
    V_H = R_H(T,B) \, B_z \, I_\mathrm{ex} \, .
    \label{eqn:V_H}
\end{equation}
With a sample placed on the device in very close proximity (height, $z\approx5$~$\mu$m) to the sensors, the previously determined sensitivity allows instead the determination of the local magnetic induction (also in the $z$-direction) from the measured Hall voltage, 
\begin{equation}
B_l = \frac{V_H} {I_\mathrm{ex} \, R_H(T,B)} \, .
\label{eqn:B_local}
\end{equation}
By measuring multiple sensors simultaneously, a spatially-resolved profile of the local $B$-field can thus be determined. The magnetisation of a sample, e.g.\ diamagnetic or ferromagnetic, creates a demagnetising field both within and around it. Therefore, the local magnetic induction, as measured by the sensors, is no longer simply the applied magnetic field but the sum of both the applied field and the sample's demagnetising field. For each sensor, we define a `local' magnetisation
\begin{equation}
    M_l = \frac{B_l}{\mu_0} - H_a \, ,
    \label{eqn:M_local}
\end{equation}
which corresponds to the sample's contribution to the local magnetic induction. Here, we use the symbol $M_l$ to differentiate between our local magnetisation and the conventional bulk magnetisation\cite{jamesFieldPenetrationSurface1997,gregorySuppressionSurfaceBarriers2001,lefebvreLocalMagnetizationFluctuations2009}.

\section{Results}

\subsection{Magnetic hysteresis}

\begin{figure*} [t]
    \centering
    \includegraphics[width=\linewidth]{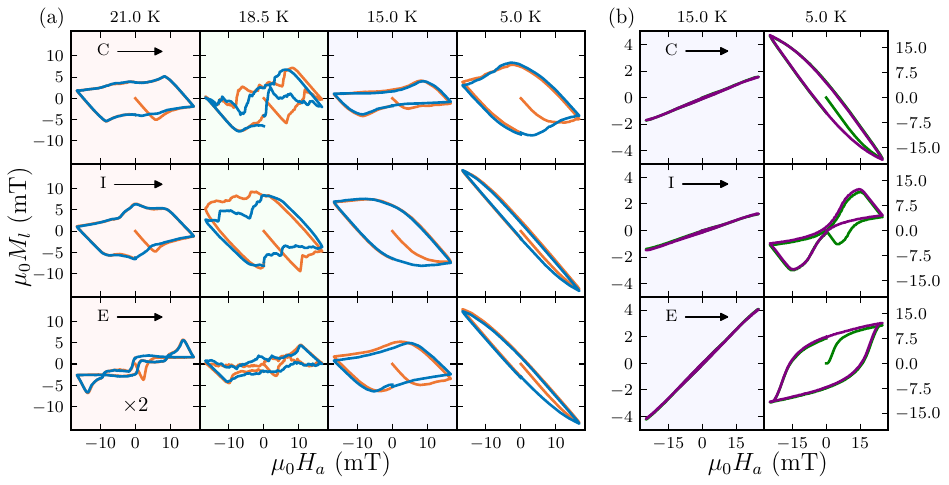}
    \caption{Hysteresis loops of local magnetisation $\mu_0M_l$ for samples (a) \#1 and (b) \#2. In each, the row corresponds to a particular sensor position - centre (C), intermediate (I) and edge (E) - beneath the sample, and the column corresponds to the given temperature. The primary and subsequent loops are indicated by the orange and blue curves in (a), and the green and purple curves in (b). The $M_l$ curve at 21.0~K for the edge sensor (E) in panel (a) has been scaled up by a factor of 2 for clarity.}
    \label{fig:2}
\end{figure*}

Local magnetisation hysteresis loops for the two samples are shown in figure \ref{fig:2}. At each temperature the samples are zero-field cooled (ZFC) from above $T_c$ to the target temperature, and $M_l$ is measured simultaneously in the three sensor positions C, I and E. The primary `virgin' loop is measured following the sequence $H_a=0\rightarrow{H_\mathrm{max}}\rightarrow-H_\mathrm{max}\rightarrow0$. A subsequent loop is also recorded immediately afterwards by repeating the same cycle of $H_a$ without warming the sample above $T_c$.

In figure \ref{fig:2}(a), at $T=21.0$~K, sample \#1 is in the superconducting-only state and the overall response is indicative of a classic type-II superconductor\cite{beanMagnetizationHardSuperconductors1962,beanMagnetizationHighFieldSuperconductors1964,brandtTypeIIsuperconductorStripCurrent1993}. At the start of the primary loop, the Meissner-like response $M_l\approx-H_a$ is evident in all three sensors, with flux penetrating at the sample edge before progressing towards the sample centre. On reaching the position of the Hall sensor, $M_l$ increases away from $-H_a$, and once flux has fully penetrated to the position of the sensor, the primary and subsequent loops are almost identical. Upon reversal of the applied field, $M_l$ increases again with slope $\mathrm{d}M_l/\mathrm{dH_a}\approx-1$ and reaches a peak, indicating the arrival of the reverse flux front. However, an anomalous peak can be seen near zero field in the intermediate sensor response, as well as some additional structure below zero field for the centre sensor. These features occur in a small window near $B_l\approx0$, potentially reflecting a field-dependent critical current $J_c(B)$. Alternatively, the cusp-like features may indicate the presence of a geometrical barrier, for which similar zero-field peaks are a characteristic signature\cite{zeldovGeometricalBarriersHighTemperature1994, jamesFieldPenetrationSurface1997, pissasEvidenceGeometricalBarriers2001}. We also note the unusual hysteretic feature in the response of the edge sensor near zero field. At this temperature the Eu$^{2+}$ moments are not yet ordered ferromagnetically but do exhibit very strong paramagnetic susceptibility\cite{grebenchukCrossoverFerromagneticSuperconductor2020}. We attribute these low field features to a contribution from the paramagnetic moments, due to the strong demagnetising fields at the edge of the sample, which saturates when the local spatial gradient exceeds the critical state (i.e.\ $dB_z/dy>\mu_0J_c$).

In figure \ref{fig:2}(a), at $T=18.5$~K$<T_\mathrm{FM}$, sample \#1 is now in the DMS - the simultaneously superconducting and ferromagnetic state characterised by narrow uniaxial stripe domains\cite{stolyarovDomainMeissnerState2018,wilcoxMagneticallyControlledVortex2025}. Here we observe highly anomalous behaviour, with the local magnetisation for all three sensors exhibits striking, nearly chaotic dependence on the applied field. Fine structure is most apparent at the centre sensor where the changes in $M_l$ appear to be indicative of the motion of clusters or chains of small numbers of vortices. The initial Meissner-like response and the slope of $M_l$ at the points of field reversal are still consistent with the type-II superconducting behaviour seen above $T_\mathrm{FM}$. However, the behaviour of $M_l$ between these regions has become almost completely irreproducible between the primary and subsequent loops. Additionally, flux appears to penetrate sooner at both the centre and edge sensors than at the intermediate sensor, suggesting an anomalous spatial variation in the flux density. 

The DMS only appears in a small window $\Delta{T}\sim1$~K below $T_\mathrm{FM}$, and so, on cooling further, sample \#1 enters instead into the DVS. Here the ferromagnetic domains recover their wider size due to the collapse of Meissner screening currents and become saturated with spontaneously nucleated vortices and antivortices\cite{stolyarovDomainMeissnerState2018,devizorovaTheoryMagneticDomain2019,wilcoxMagneticallyControlledVortex2025}. As shown in figure \ref{fig:2} at $T=15.0$ and $5.0$~K, the erratic behaviour of the DMS is replaced with a smooth, mostly reproducible evolution of $M_l$. The responses in this regime, which at first seem to resemble the typical type-II behaviour seen at $T>T_\mathrm{FM}$, in fact exhibit significant differences. At $T=15.0$~K, the point at which the flux front arrives is much less clear due to the smooth and gradual increase in $M_l$ from the Meissner-like response. Additionally, this increase of $M_l$ happens at a lower field at the centre of the sample than at the edge or intermediate positions. This is even more apparent at lower temperatures (e.g.\ 5.0~K), where the edge and intermediate positions show almost perfect Meissner screening and no penetration of magnetic flux, but the open hysteresis loop at the centre sensor indicates the surprising presence of magnetic flux that may also be linked to the existence of a geometrical barrier.

The response of sample \#2, in contrast, shows typical type-II superconducting behaviour and a high degree of reproducibility at all temperatures. In figure \ref{fig:2}(b), at 15.0~K, sample \#2 is in the ferromagnetic only state and the measured local magnetisation $M_l$ is due to the $c$-axis susceptibility of the Eu ferromagnetic order. Here the response is strongly linear with applied field and shows no measurable irreversibility, indicating that the ferromagnetic domain walls are very weakly pinned in this material. It can be seen that the local susceptibility as measured by the edge sensor is over a factor of two larger than the two other sensors because the strong demagnetising fields near the edge of the sensor enhance the local $H$-field.

Cooling sample \#2 below $T_c$, the response is now a superposition of ferromagnetic and superconducting behaviours. In figure \ref{fig:2}(b), at 5.0~K, the sample exhibits strong Meissner screening with no clear sign of flux penetration to the centre sensor. There is no contribution from ferromagnetism in the initial, Meissner-like part of the hysteresis loops ($M_l\approx-H_a$) because the applied field is perfectly screened by the superconductivity. However, the small, positive tilt of the loops outside of this region is due to the addition of the the linear susceptibility of the coexisting ferromagnetism. In the intermediate and edge positions, the loops open up due to the admittance of magnetic flux into the sample interior. The strong demagnetising field and close proximity of the edge sensor to the sample's edge results in a very rapid penetration of flux that quickly saturates except for a weak, linear contribution from the ferromagnetism. Reversing the applied field also sees a similar, rapid drop in $M_l$ and the same quasi-saturated behaviour.

\subsection{Flux penetration}

\begin{figure}
    \centering
    \includegraphics[width=0.5\linewidth]{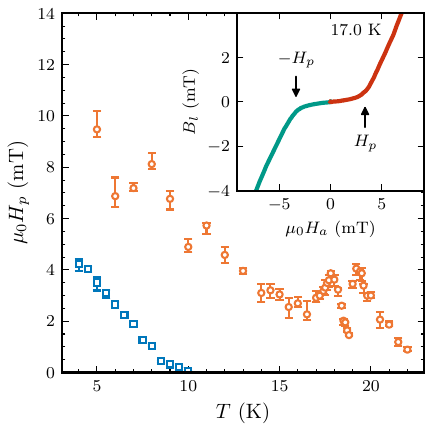} % Preprint
    \caption{Penetration field $H_p(T)$ as determined at the edge sensor for sample \#1 (orange circles) and intermediate sensor for sample \#2 (blue squares). The inset shows two example measurements at 17.0~K of $B_l(H_a)$ in sample \#1. The sample is zero-field cooled to the target temperature and the field is increased to a positive (red) or negative (green) maximum value. The penetration field $H_p$ is indicated as the field at which $B_l$ deviates from the Meissner-like response, and the error bars indicate the uncertainty in identifying $H_p$.}
    \label{fig:3}
\end{figure}

As demonstrated in figure \ref{fig:2}, measuring the local magnetic response at high spatial resolution can provide important physical insight that might otherwise be missed from global magnetic measurements that average over the entire sample. The field of first flux penetration, or penetration field $H_p$, is one such quantity that can often be misidentified in bulk probes, but is observed more easily through local magnetic measurements\cite{liangLowerCriticalField2005,lefebvreLocalMagnetizationFluctuations2009,okazakiLowerCriticalFields2009,putzkeAnomalousCriticalFields2014}. $H_p$ is the applied field at which magnetic flux penetrates the sample edge and overcomes the Meissner screening currents and, from a thermodynamic perspective, should be equivalent to the lower critical field $H_{c1}$ in a `bulk' sample\cite{tinkhamIntroductionSuperconductivity2015}. However, in practise it can often be enhanced or reduced for several reasons. The application of the field perpendicular to our platelet-shaped crystals leads to a large demagnetising factor at the sample edge, which enhances the applied field at the sample edges and reduces $H_p$ compared to $H_{c1}$ ($H_p\approx{H_{c1}}\sqrt{d/W}$). Brandt has provided a solution to this reduction of $H_p$ for certain sample geometries of pin-free superconductors\cite{brandtTypeIIsuperconductorStripCurrent1993}, and several successful uses of this approach have been reported\cite{okazakiLowerCriticalFields2009,lefebvreLocalMagnetizationFluctuations2009,putzkeAnomalousCriticalFields2014}. However, care must be taken to correctly identify $H_{c1}$ since e.g.\ strong vortex pinning / large $J_c$\cite{brandtTypeIIsuperconductorStripCurrent1993}, Bean-Livingston surface barriers\cite{beanSurfaceBarrierTypeII1964}, geometrical effects\cite{zeldovGeometricalBarriersHighTemperature1994}, or poor sensor positioning\cite{okazakiLowerCriticalFields2009} can all lead to an enhancement of $H_p$ compared to $H_{c1}$. 

In order to mitigate these various possibilities for misinterpretation, and due to the highly unusual presence of ferromagnetic domains in our circumstance, we focus first on the qualitative dependence of $H_p(T)$, as shown in figure \ref{fig:3}, taking it to be reflective in some measure of both the critical current $J_c(T)$ and lower critical field $H_{c1}(T)$. At each temperature, the sample is ZFC from above $T_c$ and $H_a$ is increased from zero to a positive maximum value. The sample is then warmed above $T_c$ and ZFC to the same target temperature, where the measurement is repeated but with a negative applied field. $H_p$ is identified empirically as the point at which the local induction $B_l$ begins to increase (or decrease) away from the Meissner-like response (figure \ref{fig:3} inset). Below $|H_p|$, the local induction is not precisely zero due to the incomplete shielding of the sensor by the sample's Meissner response, a result of the finite $z$-separation between sensor and sample\cite{putzkeAnomalousCriticalFields2014}. To compensate for any systematic offset in the applied field, e.g.\ due to the Earth's magnetic field, $H_p$ is taken as the average between the measurements in positively and negatively applied field.

For sample \#1, the edge sensor was located $\approx40~\mu$m from the sample edge, providing a clear measure of $H_p$ at that position. There are several interesting features of $H_p$(T) superimposed on a general increase as the temperature is decreased. $H_p$ peaks just above $T_\mathrm{FM}\approx19.3$~K and on cooling further, there is a very rapid suppression of $H_p$ just below $T_\mathrm{FM}$ by more than a factor of two. After this drop, $H_p$  recovers by $T\approx17.5$~K, after which there is another, smaller, decrease for around $1.5$ K. At lower temperatures the precise determination of $H_p$ becomes more challenging due to the dramatically broadened penetration of flux at the sensor location, as seen in figure \ref{fig:2}(a). Surprisingly, despite the erratic behaviour of $M_l$ in the region of the DMS, the Meissner response and corresponding $H_p$ are found to have a high degree of reproducibility between independent repeat measurements (see figure \ref{fig:5}). 

In contrast, the penetration field in sample \#2 displays a smooth, monotonic evolution with temperature on cooling below $T_c\approx12$~K. Due to the proximity of the edge sensor to the sample edge ($\sim10~\mu$m) and the finite $z$-separation between sensor and sample surface, the smooth response at the edge sensor meant it was not possible to identify a clear feature associated with $H_p$. Instead, we present $H_p$ as determined from the intermediate sensor, around $100~\mu$m from the sample edge. The values of $H_p$ are considerably lower than for sample \#1, where the sensor was closer to the sample edge, indicating both the substantially weaker superconductivity of sample \#2 and also the lower critical current, consistent with previous measurements\cite{wilcoxMagneticallyControlledVortex2025}. In this lower temperature region, where we expect the two samples to be in similar superconducting-and-ferromagnetic states, $H_p$ is rather a sharper feature in the intermediate sensor response of sample \#2 than any of the sensors of sample \#1, indicating that the different ferromagnetic domain sizes strongly influence the mechanism by which magnetic flux enters.

\subsection{Flux fronts and the critical state}

\begin{figure*}
    \centering
    \includegraphics[width=\linewidth]{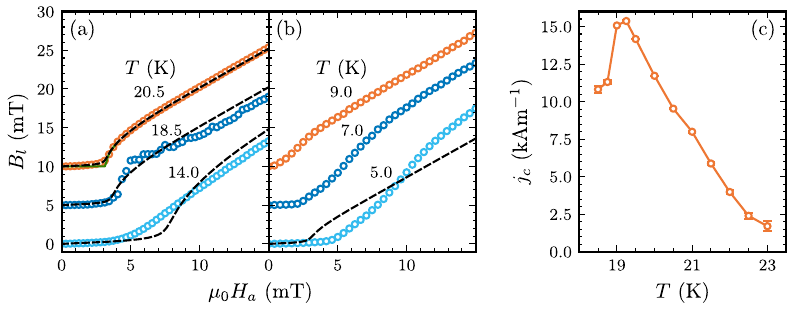}
    \caption{Local magnetic induction $B_l$ measured at different temperatures for (a) edge sensor, sample \#1, and (b) intermediate sensor, sample \#2. The hollow circles are the measured data and the dashed black lines are fits to the modified Brandt-Indenbom model (equation \ref{eq:Brandt}). The data in (a) at 18.5 and 20.5~K, and (b) at 7.0 and 9.0~K, have been offset by $+5$ and $+10$ mT for clarity. The solid green line in (a) is the expected response from the model with $z_s=0$. (c) Critical sheet current density $j_c(T)$ determined from the fits for sample \#1.} 
    \label{fig:4}
\end{figure*}

To complement our empirical analysis of $H_p(T)$, here we compare the shape of the measured flux fronts $B_l(H_a)$ to the analytical Brandt-Indenbom model\cite{brandtTypeIIsuperconductorStripCurrent1993}. The model provides a description of the supercurrent distribution $j(y)$ inside a semi-infinite thin strip of type-II superconductor due to a perpendicular applied field $H_a$ (see \textit{Supplementary Note 3} for full details). In combination with Amp\`{e}re's law, the corresponding $z$-component of the magnetic induction, $B_z$, outside of the superconductor is given to be:
\begin{equation}
\label{eq:Brandt}
    B_z(y_s, z_s) = \frac{\mu_0}{2\pi}\left( \int_{-a}^{-a} \frac{y_s-y}{(y_s-y)^2+z_s^2} j(y) \,dy \, + H_a \right).
\end{equation}
where $2a$ is the strip width, and $y_s$ and $z_s$ are the desired coordinates of the $B$-field, e.g.\ the position of a sensor, and $x,y,z$ are the directions defined in figure \ref{fig:1}.

Figure \ref{fig:4}(a) shows $B_l$ in sample \#1, as measured at the edge sensor. The sample is ZFC from above $T_c$ and the field increased from $H_a=0\rightarrow{H_\mathrm{max}}$, i.e.\ in the same manner as in figure \ref{fig:3}. At 20.5~K, sample \#1 is in the superconducting-only state. Here the flux front evolves smoothly with increasing field, and the experimental data are very well described by a fit to the model (dashed black line). Values of $y_s=108\,\mu$m and $z_s=3.6\,\mu$m were used, determined from optical micrographs, yielding a critical sheet current density $j_c= 9530$~A~m$^{-1}$. Notably, the model correctly captures the response in the Meissner-like state where $B_l$ is not precisely zero due to the finite separation between sample and sensor. The solid green line is the expected response for the same parameters but taking $z_s\rightarrow0$, i.e.\ measuring directly on the sample surface, where $B_l$ is precisely zero in the Meissner state. In the model, the arrival of the flux front leads to a sharp discontinuity in $B_l$ that is otherwise smoothed out in the experimental data due to the finite sensor-sample separation. 

On entering the DMS (figure \ref{fig:3}(a), 18.5~K), the evolution of the flux front is less well described by the same model, particularly at higher fields where the erratic behaviour is more apparent. The initial response bears a good similarity to the expected behaviour up to $\mu_0H_a\sim5$~mT, showing the smoothed-out upturn in $B_l$ in a similar manner to $T=20.5$~K. After the initial penetration, however, the flux front appears to be delayed compared to the expected behaviour for a simple type-II superconductor. This suggests that, while the form of the initial penetration of flux is not strongly influenced by the presence of the magnetic domains of the DMS, the continued propagation of that flux towards the centre of the sample is more significantly affected. 

Cooling further into the DVS (figure \ref{fig:3}(a), 14.0~K), it is clear by the very poor fit that the Brandt-Indenbom model is no longer applicable in this regime. Notably, the characteristic pronounced upturn, followed by inflection point, of the model is absent and instead, once the flux front arrives at that position, the local induction increases smoothly and steadily with $H_a$. This behaviour is very much at odds from what is expected from even the simplest critical state model\cite{beanMagnetizationHighFieldSuperconductors1964}, where the arriving front should lead to a discontinuity in $B$, and thus suggests that the mechanism of sample magnetisation has changed substantially in the DVS of sample \#1.

Figure \ref{fig:4}(b) shows similar responses for sample \#2, measured at the intermediate sensor, which is chosen over the edge sensor as providing a clearer picture of the flux penetration behaviour. In contrast to sample \#1, the responses deviate from the Brandt-Indenbom model at all temperatures. It is most clear at $T=5.0$~K that, while there is a clear point at which flux reaches the position of the sensor, the following shape exhibits an unexpected concavity (c.f.\ figure \ref{fig:4}(a), 20.5~K), not unlike in the DVS of sample \#1. Attempting to fit the model, indicated by the dashed black line, is unsuccessful and clearly underestimates the point of flux entry. This is due primarily to the difference in form of the initial flux front, but also because $B_l > \mu_0H_a$ at higher applied fields, indicating that the coexisting ferromagnetism is also contributing to the response, something completely unaccounted for in the model. In contrast, the contribution from the ferromagnetism seems to be less prominent in the response of sample \#1.

Figure \ref{fig:4}(d) shows the critical sheet current density $j_c(T)$ determined from sample \#1. Due to the unusual shape of the flux front, it was not possible to fit the model for $T<18.5$~K, and for the data points in the DMS, i.e.\ $18.5\leq{T}\leq19.5$~K, the upper limit of the fit was fixed to be $\approx50\%$ larger than the empirically derived $H_p$ values (figure \ref{fig:3}) in order not to be overly influenced by the delayed flux front at higher fields. Below $T_c$, $j_c(T)$ increases linearly with reducing temperature and reaches a peak very near $T_\mathrm{FM}$, before dropping slightly. This behaviour contrasts to $H_p(T)$ (figure \ref{fig:3}) which saw a much more dramatic decrease over the same temperature range. Instead, $j_c(T)$ is more consistent with recently reported bulk magnetic measurements of a sample with the same nominal composition\cite{wilcoxMagneticallyControlledVortex2025}. It is clear that while the onset of ferromagnetism, and the appearance of the DMS, is inherently detrimental to superconductivity, the corresponding effect on the critical current is less pronounced.

\subsection{Magnetic irreversibility in the domain Meissner state}

\begin{figure*}
    \centering
    \includegraphics[width=\linewidth]{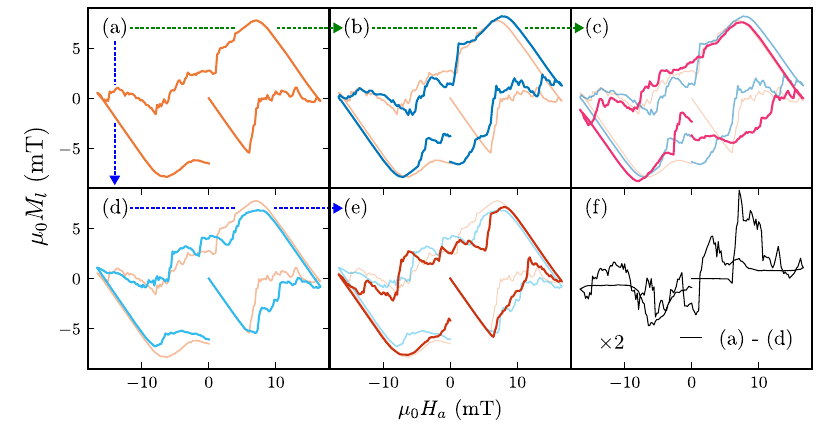} 
    \caption{Local magnetisation hysteresis loops measured in sample \#1 at 18.5~K using the centre sensor. (a)-(c): After ZFC from above $T_c$, the orange (a), blue (b) and pink (c) curves are the first, second and third loops measured in immediate succession. The previous loops to (b) and (c) are shown with increasing transparency. (d), (e): The light blue and red curves are independent repeats with the sample ZFC from above $T_c$ between each loop. The curve from (a) is shown as transparent in (d), and similarly (a) and (d) are shown as transparent in (e). (f): The difference between the response in (a) and in (d), i.e.\ $\Delta M_l = M_l^{(a)} - M_l^{(d)}$. The data have been expanded by a factor of 2 for clarity. The green and blue arrows indicate the sequence of successive measurements.}    
    \label{fig:5}
\end{figure*}

From figure \ref{fig:2}(a), it is clear that the onset of the DMS causes a significant change in the behaviour of the magnetic irreversibility in sample \#1 due to the presence of the finely-spaced ferromagnetic stripe domains characteristic of the DMS. Most striking are the large fluctuations in magnetisation that are not reproducible between successive loops; here we explore this behaviour in more detail. In figure \ref{fig:5}(a)-(c), the sample is ZFC from above $T_c$ to 18.5~K and a sequence of three successive hysteresis loops are recorded (sequence indicated by dashed green arrows). There is some broad structure to the loops that is reproducible between successive measurements: $M_l$ increases smoothly on reversal of $H_a$ at $H_\mathrm{max}$, and the interior `size' of the loops are approximately the same. After the point of field reversal at $H_\mathrm{max}$, $\mathrm{d}M_l/\mathrm{d}H_a\approx-1$ until around $H_a\sim6$-7~mT, where a peak in $M_l$ is reached. Beyond this peak, $M_l$ decreases smoothly at first before transitioning into large, erratic fluctuations as $H_a$ is reduced towards $-H_\mathrm{max}$. It is in this upper portion of the loops, i.e.\ between the peak and $-H_\mathrm{max}$, as well as the equivalent lower portion, where the precise evolution of $M_l$ becomes considerably less reproducible. 

Similarly, in figure \ref{fig:5}(a, d \& e) (indicated by dashed blue arrows), repeat ZFC of the sample from above $T_c$ (to reset the magnetic state) results in behaviour that shows some larger-scale reproducibility but significant variations on a finer scale. Such variability is made clearer by plotting the difference between independent, repeat measurements, $\Delta M_l = M_l^{(a)} - M_l^{(d)}$, as shown in figure \ref{fig:5}(f). The initial flat line where $\Delta M_l = 0$ corresponds to the same Meissner screening in both measurements. Similar flat lines are also evident at the point of field reversal. However, between these flat portions, $\Delta M_l$ displays large and seemingly random fluctuations. We attribute this variability of the fine-scale structure of $M_l$ to the motion of superconducting vortices through the narrowly-pitched ferromagnetic domains of the DMS, and the formation of vortex polarons therein\cite{wilcoxMagneticallyControlledVortex2025}. The lack of repeatability can be understood as being due to the different realisations of the local magnetic domain structure after each cool down, as well as after each modification and restructuring of the domains by the penetrating flux fronts. However, some quasi-reproducible features in these fluctuating sections do also suggest that there are some repeatable flux-domain processes taking place, e.g.\ the drop in $M_l$ in the upper branch at $H_a\approx1$ mT, followed by a broad hump, as shown in figure \ref{fig:5}(b). 

 \begin{figure}
     \centering
     \includegraphics[width=0.5\linewidth]{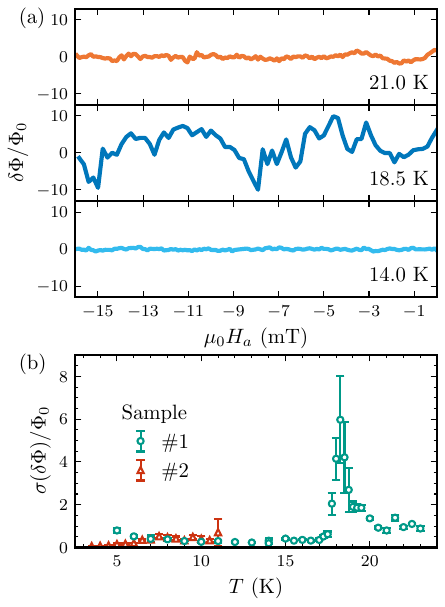} % Pre-print 
     \caption{(a) Sections of hysteresis loops from sample \#1 in the upper, $H_a$-decreasing branch at three temperatures using the centre sensor. The coloured, solid lines are the data after subtracting a 4th-order polynomial, in terms of the flux quantum $\Phi_0$. (b) The size of the characteristic fluctuations $\sigma(\delta\Phi)$ as a function of temperature. The error bars represent variations in the characteristic fluctuations from multiple hysteresis loops.}
     \label{fig:6}
 \end{figure}

To characterise the size and temperature dependence of this erratic behaviour, we perform a fluctuation analysis\cite{lefebvreLocalMagnetizationFluctuations2009} of the magnetisation in the upper and lower branches. This is done by subtracting a smooth background from sections of hysteresis loops to isolate the fine structure; a 4th-order polynomial was found to be a good approximation without over-fitting the data. Figure \ref{fig:6}(a) shows the result of this background subtraction for three temperatures in the upper, $H_a$-descending portion of the loops. The data have been presented in terms of the magnetic flux $\Phi=B_lA$, where $A=25\mu$m$^2$ is the active area of the Hall sensor, and $\delta\Phi$ are the data after subtracting the background. It is clear how much larger the variations in flux are when in the DMS, at 18.5~K, than either the superconducting-only state or the DVS. The large fluctuations of up to tens of flux quanta, $\Phi_0$, suggest highly collective and irregular motion of chains/clusters composed of multiple superconducting vortices.

Figure \ref{fig:6}(b) shows the size of these characteristic fluctuations as a function of temperature for samples \#1 and \#2. Here, $\sigma(\delta\Phi)$ is the average of the standard deviations of $\delta\Phi$ derived from multiple hysteresis loops, with the error bars representing the variations in the set of values. In sample \#1 above $T_\mathrm{FM}$, $\sigma(\delta\Phi)\sim1~\Phi_0$, demonstrating our ability to resolve the entrance and exit of single flux quanta from the active area of our sensor. Furthermore, in the superconducting-only state, the flux are essentially uncorrelated, indicating the intrinsically weak-pinning nature of isovalent P-doping in our samples\cite{vanderbeekQuasiparticleScatteringInduced2010}.

Within the region of the DMS, it is very clear that the large fluctuations onset just below $T_\mathrm{FM}\approx19.3$~K and increase rapidly up to a maximum of $\sim6~\Phi_0$ at $T\approx18.25$~K. Notably, this peak occurs at a slightly lower temperature than the minimum in $H_p$ (figure \ref{fig:3}). As the temperature is reduced further, and the sample transitions into the DVS, the size of the fluctuations now decrease rapidly. By $T\approx16.5$~K, they are reduced to $<1~\Phi_0$ and remain very small down to the lowest temperatures. The size of the fluctuations in sample \#2 are also similarly small as in the DVS of sample \#1, reducing even further to become nearly unresolvable by $T=3.0$~K.

\section{Discussion}

As well as bulk pinning of vortices, there are alternative mechanisms that could drive magnetic irreversibility in EuFe$_2$(As$_{1-x}$P$_x$)$_2$. Firstly, the extremely small coercivity of sample \#2 in the ferromagnetic-only state (c.f.\ figure \ref{fig:2}(b)) indicates that the material is an exceptionally soft ferromagnet with very weak domain wall pinning, and a detailed analysis of magnetic behaviour at optimal composition\cite{wilcoxMagneticallyControlledVortex2025} implies sample \#1  has very similar properties. Therefore, we argue the origin of magnetic hysteresis in both samples is due to the irreversible motion of superconducting vortices, whether in the superconducting-only or any of the coexistence states, and not due to any intrinsic ferromagnetic domain pinning. 

Additional sources of irreversibility could come from Bean-Livingston-type surface barriers\cite{beanSurfaceBarrierTypeII1964} or, as discussed previously, barriers of geometrical origin\cite{zeldovGeometricalBarriersHighTemperature1994}. In the case of a Bean-Livingston surface barrier, this can lead to phenomena such as anomalous magnetisation loops in the vicinity of $T_c$ where the descending branch sees a magnetisation close to zero\cite{konczykowskiEvidenceSurfaceBarriers1991}, or distinct upturns in the penetration field at low temperatures\cite{burlachkovExplanationLowtemperatureBehavior1992}. The Bean-Livingston barrier is driven in part by the image force a vortex experiences in the vicinity of the sample surface, requiring the surface to appear smooth and without irregularities on the length-scale of the penetration depth $\lambda$. Thus, it is possible for this kind of barrier to manifest in some materials, particularly high temperature superconductors, near $T_c$ where $\lambda$ is much larger, and for the effects to disappear at lower temperatures where $\lambda$ reduces below the length-scale of the surface irregularities. While it is difficult to rule out entirely, there are no particularly obvious signs that the irreversibility is driven by a Bean-Livingston surface barrier in either the superconducting-only state of sample \#1 nor in the coexisting regions of both of the samples. Furthermore, the presence of ferromagnetic domains must surely modify the underlying image force behaviour.

In contrast, the zero-field cusps in the hysteresis loops of sample \#1 in the superconducting-only state (figure \ref{fig:2}(a), $T=21.0$ K) are likely signs of a geometrical barrier in this sample\cite{zeldovGeometricalBarriersHighTemperature1994}. Such geometrical barriers can occur in thin, flat superconductors with relatively weak bulk pinning ($J_c$) and strong edge currents ($J_E$), and are typically more evident nearer to $T_c$ when flux pinning is naturally weaker compared to edge currents. In the limit of zero bulk pinning, the irreversible state due to the geometrical barrier results in a non-Bean-like flux dome in the centre of the sample that grows outwards with increasing field $H_a$. With small amounts of bulk pinning, the dome transforms instead into an annulus that grows both towards the centre as well as to the sample edges. Therefore, the relatively larger size of the loop at the intermediate sensor compared to the centre sensor can be interpreted as evidence of a geometrical barrier at this temperature, though there is evidently some modest bulk pinning as well.

At lower temperatures, there are further signs for an anomalous spatial distribution of flux. This is particularly clear in figure \ref{fig:2}(a) at $T=5.0$~K, where flux initially penetrates at the centre of the crystal before reaching the intermediate and edge sensor elements (the afore-described flux dome). However, there are no longer zero-field cusps evident in any of the hysteresis loops that were seen closer to $T_c$. While localisation of flux at the centre of the sample may suggest the persistence of a geometrical barrier to lower temperatures, the increase in $J_c$ as temperature is decreased would tend to see the associated effects become weaker, not stronger\cite{jamesFieldPenetrationSurface1997}, and for the annulus to move instead towards the sample edges\cite{zeldovGeometricalBarriersHighTemperature1994}. Further, the existence of ferromagnetic domains saturated with spontaneously nucleated vortices and antivortices in the ZFC state of the DVS presents a substantial deviation from many of the assumptions that underpin the theory describing the geometrical barrier. This behaviour is also conspicuously absent in sample \#2, such that it does not appear to be intrinsically linked to the presence of the underlying ferromagnetic domain structure. We believe this unusual irreversibility to exist in addition to more conventional bulk pinning, and further work is required to understand the origin of this behaviour.

While the thickness of sample \#2 is approximately three times that of sample \#1, both samples are very much taller than the characteristic length-scales of both the superconductivity (the magnetic penetration depth and the coherence length) and the ferromagnetic domain structure (the domain width and domain period). Furthermore, the lateral in-plane dimensions of both samples are much larger than their thicknesses, resulting in comparable demagnetising factors of $N\approx0.97$ (sample \#1) and $N\approx0.91$ (sample \#2)\cite{prozorovEffectiveDemagnetizingFactors2018}. Thus we are confident that both samples should give rise to similar magnetisation phenomena, and that any difference between the two cannot be due to only their relative sizes.

As such, the distinct \textit{difference} in the magnetic behaviours in the two samples highlights the acute sensitivity of the irreversibility,driven by bulk pinning, to the specific parameters of the domain structure. At the nominally optimal composition of $x\approx0.21$, sample \#1 displays distinct magnetic irreversibility behaviours corresponding to the different coexisting magnetic states. Type-II superconducting behaviour is evident above $T_\mathrm{FM}$, while strongly fluctuating and irreproducible magnetic behaviour appears below $T_\mathrm{FM}$, precisely in the expected narrow region of temperature of the DMS. The fluctuations disappear as the temperature is reduced, evolving into a smoother and more repeatable behaviour precisely at the temperature where the DVS is expected to manifest. In contrast, sample \#2, with $T_c<T_\mathrm{FM}$, only displays a single type of magnetic irreversibility reflecting the fact that it likely immediately enters into a DVS-like state upon cooling below $T_c$.

In sample \#1, just below $T_\mathrm{FM}$, the appearance of the ferromagnetic domain structure of the DMS causes spontaneous Meissner currents to screen the magnetisation of the domains, increasing the kinetic energy of the superconducting state and suppressing the superconducting order parameter. The corresponding loss of superfluid density means that the penetrating flux front can overcome the screening currents more easily, resulting in the sharp drop of $H_p$ at the same temperature. Recent reports of the critical current found instead a peak at the same temperature\cite{wilcoxMagneticallyControlledVortex2025}, indicating that our measurements of $H_p(T)$ are more sensitive to $H_{c1}(T)$ than $J_c(T)$. Within the DMS, the superconducting order parameter is most strongly suppressed at the domain walls where the screening currents are largest\cite{stolyarovDomainMeissnerState2018}. As the temperature is reduced, the domains grow rapidly, reducing the domain wall density and leading to a recovery in the superfluid density and $H_p$. It is also expected that the appearance of the DMS is suppressed to a temperature slightly below $T_\mathrm{FM}$ due to electromagnetic screening by the superconductivity\cite{devizorovaTheoryMagneticDomain2019}. The peak in the magnetic fluctuations at $T\approx18.25$~K ($<T_\mathrm{FM}$) in figure \ref{fig:6}(b) demonstrates the requirement for the ferromagnetic order parameter to continue growing before the domain structure is sufficiently well established to enable the collective vortex motion that we associate with the fluctuations to occur.

The reduction seen in $H_p$ between $16.5\lesssim{T}\lesssim17.5$ K is likely due to the sample's smooth transition from the DMS to the DVS. As the temperature is reduced, the fractional occupation of the sample by regions of DVS increases until eventually reaching saturation\cite{stolyarovDomainMeissnerState2018,wilcoxMagneticallyControlledVortex2025}. In the areas of DVS, the superconducting order parameter is further suppressed due to the spontaneous nucleation of vortices and antivortices, although the screening Meissner currents of the DMS disappear at the same time. In this temperature window, the ferromagnetic order parameter is growing rapidly, increasing the magnetisation of the stripe domains and therefore the flux density. As the sample transitions all the way to the full DVS at $T\approx16.5$~K, the reduction in $H_p$ ends quite abruptly, and instead it starts to slowly increase again towards zero temperature. The transition from large to small fluctuations in figure \ref{fig:6}(b) occurs very close to the temperature where we see a reversal in the slope of $H_p(T)$, further emphasising these as clear hallmarks of the DMS to DVS transition.

The nearly linear increase in $H_p(T)$ that we observe at low temperatures could be linked to a thermally-activated vortex-antivortex annihilation mechanism by which the DVS becomes magnetised\cite{wilcoxMagneticallyControlledVortex2025}. In zero field, the DVS consists of domains of alternating up and down magnetisation which are filled equally with vortices and antivortices, respectively. In an applied field, up (down) domains will grow (shrink) to minimise the magnetostatic energy\cite{kooyExperimentalTheoretical1960}. To do so requires the domain walls to move, except here the domains are filled with mutually-repulsive superconducting vortices that cannot reconfigure freely. Instead, a Bean-Livingston barrier at domain walls must be overcome for vortex-antivortex annihilation to occur and allow the reshaping of domain sizes. In addition, vortices entering from outside of the sample will be met with a sea of spontaneous vortices, greatly modifying their ability to redistribute within the individual domains due to the saturation of available intrinsic pinning sites. The observation that $\sigma(\delta\Phi)<1~\Phi_0$ in both samples at low temperatures (figure \ref{fig:6}(b)) strongly suggests that flux entry/exit into the sensor area is no longer well characterised in terms of individual fluxons, rather vortex motion is strongly screened by the sea of pre-existing vortices.

With $T_c\approx12$~K, we estimate $x\approx0.28$ for sample \#2, placing it far into the over-doped region of the phase diagram (figure \ref{fig:1}(a)). While we do not have any direct visualisation of the underlying ferromagnetic state in this sample, measurements of magnetic susceptibility indicate it has almost exactly the same Curie temperature $T_\mathrm{FM}\approx19.3$~K as sample \#1. Thus, it seems likely that uniaxial stripe domains are universal to the Eu-ferromagnetism portion of the phase diagram. Substitution of P for As in the crystal structure causes a reduction in the $c$-axis lattice parameter and brings the Eu atoms closer\cite{marchandTernaryLanthanoidtransition1978,jiangSuperconductivity302009}, modifying the interlayer ferromagnetic coupling and magnetic anisotropy\cite{zapfVaryingEu2011}, thus influencing the sizes of the domains and domain walls. An MFM study on an over-doped sample, with $x=0.25$ and $T_c\approx18.4$~K~$<T_\mathrm{FM}$, also found a uniaxial stripe domain structure, but with a domain width $\approx50\%$ wider than at optimal doping\cite{grebenchukCrossoverFerromagneticSuperconductor2020}. Notably, on cooling below $T_c$ at $H = 0$, this sample was observed to enter directly into a densely-populated vortex-antivortex state similar to the DVS found at optimal composition\cite{stolyarovDomainMeissnerState2018}. Consequently, we expect the domains in our $x\approx0.28$ sample (\#2) to be even wider than those observed at $x\approx0.25$. Additionally, since the ferromagnetic state is already fully developed as we cool this sample below $T_c$, we should not observe the DMS. Rather, we expect the domains to generate spontaneous vortices and antivortices as soon as the sample becomes superconducting, comparable to the DVS at optimal doping. Hence, while both our samples are expected to exist in similar DVS-like magnetic states at low temperatures, it is surprising to find they differ significantly in how the magnetic irreversibility and the critical state behave. 

To address this disparity, we speculate that there are novel processes in sample \#1, enabled by narrower domain widths, that are otherwise absent in sample \#2. We suggest that the magnitude of the ferromagnetic domain widths play crucial, but very distinct, roles in the DMS and DVS phases. In the DMS, the vortex polaron binding energy depends strongly on the ratio $\lambda/w_D$, and the very narrow domains observed in sample \#1 in this regime are responsible for the enhanced vortex creep activation energy and vortex clustering\cite{wilcoxMagneticallyControlledVortex2025}. In contrast, in the DVS and DVS-like states, magnetisation is expected to involve vortex-antivortex annihilation at domain walls. Therefore, the higher the wall density, $\propto1/w_D$, the more rapid this process will be, which in turn qualitatively explains why the magnetic reversal in sample \#2, with a much lower domain wall density, is strikingly different from \#1 at low temperatures.

The validity of the thin-film critical state model of Brandt and Indenbom breaks down in the low temperature DVS-like state of both samples (figure \ref{fig:4}), suggesting that one or more of the model assumptions is not satisfied in this regime. One obvious shortcoming is linked to the fact that the shapes of our samples are closer to square, and not the infinite strip assumed in the theory. This can result in flux penetrating to a given sensor from more than one direction, rather than just from the closest edge as assumed in the model. However, the fact that good agreement is obtained with theory at higher temperatures, in both the DMS and purely superconducting regimes of sample \#1, suggests that this is not the critical factor. Instead, we believe the most important factor is the relatively wide stripe domains underlying the DVS state which breaks the model's assumption of uniform and spatially-homogeneous flux pinning. In practice, vortices will be guided to penetrate at the edges of the samples \textit{along} domains with parallel magnetisation leading to a natural modulation of the shape of the flux front. Similar types of edge disorder have been shown to generate strong flux front roughening\cite{geahelEdgeContamination2017}, resulting in the strong rounding of the measured penetration field since different parts of the flux front cross into the sensor at different applied fields.

\section{Conclusion}

In conclusion, we report a spatially-resolved study of the irreversible magnetisation in the robust ferromagnetic superconductor EuFe$_2$(As$_{1-x}$P$_x$)$_2$. We reveal a rich variety of novel magnetic behaviours that are strongly correlated with underlying ferromagnetic domain structures. In the superconducting-only state, zero-field cusps in magnetic hysteresis loops suggest an energy barrier of geometrical origin. In the DMS, the narrowly-spaced domains give rise to highly erratic and irreproducible fluctuations in the magnetisation that we attribute to the collective motion of clusters containing multiple flux quanta, stabilised by the formation of vortex polarons. In contrast, the saturated, wider domains of the DVS lead to a markedly smoother evolution of the irreversible state whereby the penetrating flux front is roughened by the presence of the domains, spontaneously breaking one of the assumptions of the critical state model. Our results indicate that the mechanism governing irreversibility is strongly influenced by the precise nature of the underlying ferromagnetic domains, being very sensitive to the specific material parameters of EuFe$_2$(As$_{1-x}$P$_x$)$_2$. We propose that manipulation of the magnetic domain structure into loops, bubbles, or other shapes via additional external controls, e.g.\ in-plane magnetic fields or mechanical strain, could give rise to further novel vortex-domain processes and related magnetic behaviours.

\section{Acknowledgements}

J.A.W.\ and S.J.B.\ acknowledge support from the Engineering and Physical Sciences Research Council (EPSRC) in the United Kingdom under Grant No.\ EP/X015033/1. W.R.F. acknowledges receipt of a PhD studentship supported by the University of Bath EPSRC Doctoral Landscape Award.

W.R.F., J.A.W.\ and S.J.B.\ initiated this work. T.R.\, I.V.\ and T.T.\ grew the samples. W.R.F.\ and J.A.W.\ performed the measurements. J.A.W., W.R.F.\ and S.J.B.\ prepared the manuscript with input from all authors.

\section{Conflict of interest}

The authors declare that they have no known competing financial interests or personal relationships that could have appeared to influence the work reported in this paper.

\section{Data availability statement}

The data that support the findings of this study are openly available in the University of Bath Research Data Archive at (DOI to be inserted here)\cite{fernDatasetBreakdownCritical2025}.

\section{References}

\bibliographystyle{unsrt}
\bibliography{main-bib}

\end{document}

\title{Supplementary information for\\``Breakdown of the critical state in the ferromagnetic superconductor EuFe$_2$(As$_{1-x}$P$_x$)$_2$''}

\author{William Robert Fern}
\affiliation{Department of Physics, University of Bath, Claverton Down, Bath, BA2 7AY, United Kingdom}

\author{Joseph Alec Wilcox}
\email{Corresponding author: jaw73@bath.ac.uk}
\affiliation{Department of Physics, University of Bath, Claverton Down, Bath, BA2 7AY, United Kingdom}

\author{Tong Ren}
\affiliation{Department of Applied Physics, The University of Tokyo, 7-3-1 Hongo, Bunkyo-ku, Tokyo 113-8565, Japan}

\author{Ivan Veshchunov}
\affiliation{Department of Applied Physics, The University of Tokyo, 7-3-1 Hongo, Bunkyo-ku, Tokyo 113-8565, Japan}

\author{Tsuyoshi Tamegai}
\affiliation{Department of Applied Physics, The University of Tokyo, 7-3-1 Hongo, Bunkyo-ku, Tokyo 113-8565, Japan}

\author{Simon John Bending}
\affiliation{Department of Physics, University of Bath, Claverton Down, Bath, BA2 7AY, United Kingdom}

\maketitle

\section*{Supplementary Note 1: EuFe$_2$(As$_{1-x}$P$_x$)$_2$Crystal Growth}

Samples of EuFe$_2$(As$_{1-x}$P$_x$)$_2$ are grown using a self-flux method\cite{wilcoxMagneticallyControlledVortex2025}. Powders of FeAs, FeP, and Eu (99.99\%) are mixed in stoichiometric amounts and loaded into alumina crucibles. The crucibles are then placed and sealed in stainless steel tubes under Ar atmosphere. The tubes are heated under N$_2$ atmosphere to $\geq1300~^\circ$C and held there for 12 hours. Next, the crucibles are cooled slowly to $1050~^\circ$C at $2~^\circ$C per hour, after which they are cooled naturally to room temperature. This growth method produces platelet-shaped single crystals with the shortest dimension corresponding to the crystal $c$-axis.

\section*{Supplementary Note 2: Calibration of Hall sensor array devices}

At low temperatures, the Hall response of the 2DEG sensors is well-described by a very weakly non-linear (quadratic) function:
\begin{equation}
    R_H (T, B) = R_{H,0}(T)(1+\beta B^2) \, .
    \label{eqn:R_H_beta}
\end{equation}
Here, $R_{H,0}(T)$ is a temperature dependent quantity that describes the nearly-linear low-field ($\sim$~few~mT) response of each Hall sensor, i.e. $V_H\propto{B}$. At the larger extremes of the applied field ($H_\mathrm{max}\simeq\pm\,25$~mT), the deviation of the response from linear is more apparent and is characterised by the $\beta$ term ($\beta\approx8\times10^{-5}$~mT$^{-2}$). This was found to be nearly universal amongst the various sensors and very weakly temperature dependent, reflecting the small, intrinsic non-linearity of the response of the 2DEG to larger magnetic fields.

As described in the Methods section of the main text, the reference sensor is located far away from the superconducting samples in order to provide an in-situ determination of the local magnetic induction. Measurements are made at various temperatures in order to determine the parameters $R_{H,0}(T)$ and $\beta(T)$ for the reference sensor, since $B_\mathrm{Ref} = \mu_0H_a$. Using these temperature dependent parameters, the local magnetic induction can then be determined for any other sensor (e.g.\ C, I or E) by scaling at a fixed temperature $T_0 > T_c$ (typically taken to be $T_0 = 80$ K):
\begin{equation}
    R_H^{\mathrm{C,I,E}}(T,B) = \frac{R_{H,0}^{\mathrm{C,I,E}}(T_0)}{R_{H,0}^{\mathrm{Ref}}(T_0)}R_H^{\mathrm{Ref}}(T, B) \, .
    \label{eqn:scaling}
\end{equation}

\begin{figure*}
    \centering
    \includegraphics[width=0.8\linewidth]{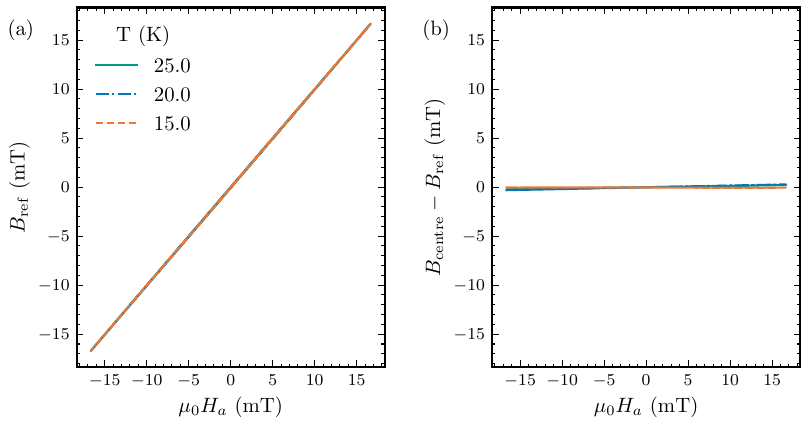}
    \caption{(a) The measured magnetic induction of the reference sensor with no sample on the device. (b) The difference in the measured magnetic induction of the reference sensor and the inferred magnetic induction of the centre sensor.}
    \label{fig:SI_1}
\end{figure*}

\begin{figure*}
    \centering
    \includegraphics[width=0.8\linewidth]{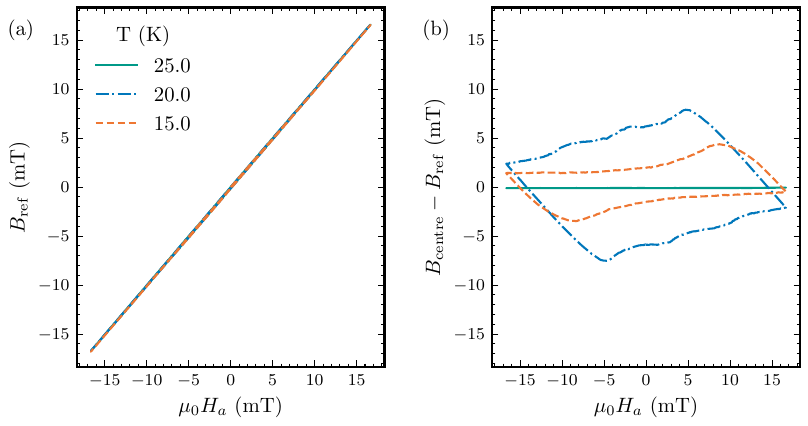}
    \caption{(a) The measured magnetic induction of the reference sensor with sample \#1 mounted on the device. (b) The difference in the measured magnetic induction of the reference sensor and the inferred magnetic induction of the centre sensor.}
    \label{fig:SI_2}
\end{figure*}

Supplementary Figure \ref{fig:SI_1}(a) shows representative measurements of $B_l$ at the position of the reference sensor when no sample is present on the device, and \ref{fig:SI_1}(b) shows the difference between the \textit{measured} magnetic induction at the reference sensor and the \textit{inferred} magnetic induction at the centre sensor based on equation \ref{eqn:scaling} and $V_H(H_a)$ as measured for the centre sensor. The difference between the two is very close to zero, though there is a small, smoothly field-dependent component that is not completely accounted for. Supplementary Figure \ref{fig:SI_2} shows a similar set of measurements, though this time with sample \#1 mounted on the device. It is very clear in 2(b) that small residual component of the calibration procedure is very much smaller than the response from the sample in the superconducting state ($T_c \approx 24$ K), and above $T_c$ the difference between the two is almost negligible. Furthermore, the smoothly-varying nature of the residual component means it cannot contribute any form of hysteretic or non-monotonic behaviour.

\section*{Supplementary Note 3: Brandt-Indembom model of type-II superconductor strip}

\begin{figure} [h]
    \centering
    \includegraphics[width=0.4\linewidth]{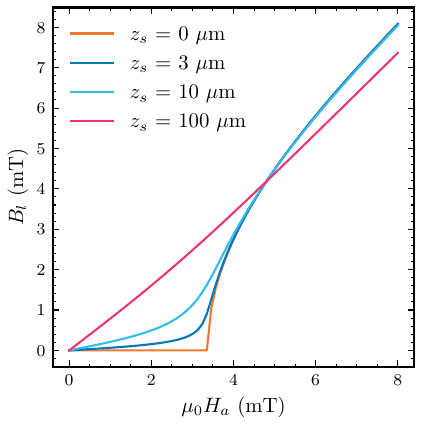}
    \caption{Calculated local magnetic induction $B_l$ at different heights $z_s$ below the sample surface utilising representative values for sample \#1: $y_s = 108\,\mu$m, $j_c = 10$~kA~m$^{-1}$ and $2a = 297$~$\mu$m.}
    \label{fig:SI_3}
\end{figure}

The model assumes a semi-infinite strip to $\pm\infty$ in the $x$-direction, occupying the space $|z|<d/2$ and $|y|<a$, where $d$ corresponds to the strip's thickness and $2a$ to its width. Here our samples are mounted on the Hall sensor array device such that the sequence of sensors measures the penetration of flux along $y$ at the midpoint of the sample's long edge. We take this as being an approximation of the infinite edge in the Brandt-Indenbom model\cite{brandtTypeIIsuperconductorStripCurrent1993}, and thus $w\approx2a$ and $d\approx{t}$. In the case of the `virgin' state, i.e.\ ZFC and $H_a$ increasing, the local magnetic field in the $z$-direction at a point in space ($y_s,z_s$) outside the strip, i.e.\ at a sensor's position, is given by
\begin{equation}
\label{eqn:Hz}
    H_z(y_s, z_s) = \frac{1}{2\pi}\ \int_{-a}^{-a} \frac{y_s-y}{(y_s-y)^2+z_s^2} j(y) \,dy \, + H_a .
\end{equation}
The quantity $j(y)~\simeq~J(y,z)d$ is the sheet current density along the sample's width, and, in an applied field $H_a$ parallel to the $z$-direction, is given by
\begin{equation}
    j(y) = 
    \begin{cases}
        \frac{2j_c}{\pi}\arctan{\frac{cy}{\sqrt{b^2-y^2}}}&|y|<b\\
        \frac{j_c y}{|y|}&b<|y|<a\,,
    \end{cases}
    \label{eqn:jy}
\end{equation}
where $j_c$ is the critical sheet current density, $H_c=j_c/\pi$,
\begin{equation}
   b = \frac{a}{\cosh{\frac{H_a}{H_c}}}\, ,\qquad
   \mathrm{and}\qquad
   c = \tanh{\frac{H_a}{H_c}}
   \label{eqn:bc}
\end{equation}

\begin{figure*}
    \centering
    \includegraphics[width=0.8\linewidth]{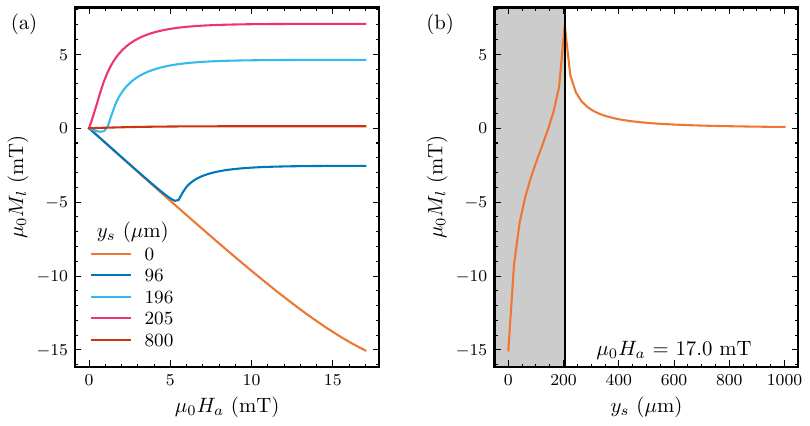}
    \caption{Calculated local magnetisation $M_l$ (a) as a function of applied field $H_a$ at different lateral positions $y_s$ utilising representative values for sample \#2: $z=3\,\mu$m, $j_c = 10$~kA~m$^{-1}$ and $2a = 410$~$\mu$m, and (b) as a function of lateral position for fixed applied field using same values as in (a). The shaded region in (b) represents values of $y_s$ which are directly beneath the sample.}
    \label{fig:SI_4}
\end{figure*}

Supplementary Figure \ref{fig:SI_3} shows the calculated local magnetic induction using equation \ref{eqn:Hz} and representative values of sample \#1 for different heights $z$ below (or, equivalently, above) the sample strip. Here, $B_l \equiv \mu_0 H_z$. On the surface of the sample ($z=0$) the applied field is completely screened by the sample's demagnetising field, i.e.\ $B_l = 0$, which is what we term the `Meissner-like' screening. The arrival of the flux front at the $y$-position of the sensor at $\mu_0H_a \approx 3.5$ mT sees a sudden, discontinuous increase in the local induction with a shape characterised by $j_c$. Repeating this calculation for increasing values of $z_s$ sees, firstly, that the Meissner-like response no longer results in precisely $B_l = 0$ at the small distance outside of the sample. Additionally, the arrival of the flux front leads to an increase in $B_l$ at a lower value of $H_a$ than when $z_s=0$. At higher fields the responses at $z=0$ and $z>0$ do become nearly coincident. At a distance far beneath the sample ($z_s=100\,\mu$m in Supplementary Figure \ref{fig:SI_3}) the influence of the sample's demagnetising field is almost negligible and $B_l \approx \mu_0 H_a$.

Supplementary Figure \ref{fig:SI_4}(a) shows calculations of the local magnetisation ($M_l = B_l/\mu_0 - H_a$) as a function of applied field. Here, representative values of sample \#2 are used, and the expected response is shown at a number of lateral positions with fixed $z_s = 3\,\mu$m. The positions $y_s=0,96,196\,\mu$m correspond to the approximate positions of the centre, intermediate and edge sensors (as described in the main text, Methods section). The position $y_s=205\,\mu$m corresponds to the very edge of the sample (having a width $2a=410\,\mu$m), and $y_s=800\,\mu$m corresponds to the approximate position of the reference sensor. 

Beneath the centre of the sample, the applied field is well screened and $M_l \approx -H_a$ up until $\mu_0H_a\approx15$ mT where there is a very gentle flattening of the response. At the position of the intermediate sensor, the Meissner-like response is evident at first until the flux front arrives much earlier at $\mu_0 H_a\approx 5$ mT, with $M_l$ displaying a small increase but remaining negative up to the maximum field. At the edge of the sample ($y_s=196$ and $205\,\mu$m), flux penetrates very quickly and the local magnetisation is in fact positive due to the demagnetising field coming from the sample. The effect of the demagnetising field is more clear in Supplementary Figure \ref{fig:SI_4}(b), which shows the expected response as a function of lateral position $y_s$ and fixed values of $z_s$ and $H_a$. In the space immediately surrounding the sample, there is a very strong effect coming from the sample's demagnetising field resulting in an enhancement of the local induction, indicated by the fact that $M_l > 0$. However, this effect drops rapidly and, at the approximate position of the reference sensor ($y_s=800\,\mu$m), the local magnetisation is very close to zero, or, in other words, the local induction is simply due to the applied field, $B_l = \mu_0H_a$. More precisely, at $y_s=800\,\mu$m (as in the figure) $M_l/H_a\approx 0.5\%$.

\bibliographystyle{unsrt}
\bibliography{SI-bibliography}